# Local-Testability and Self-Correctability of q-ary Sparse Linear Codes

Widad Machmouchi
Computer Science and Engineering
University of Washington
Seattle, WA 98195-2350
widad@cs.washington.edu

**Abstract**

We prove that q-ary sparse codes with small bias are self-correctable and locally testable. We generalize a result of Kaufman and Sudan [3] that proves the local testability and correctability of binary sparse codes with small bias. We use properties of q-ary Krawtchouk polynomials and the McWilliams identity -that relates the weight distribution of a code to the weight distribution of its dual- to derive bounds on the error probability of the randomized tester and self-corrector we are analyzing.

## 1 Introduction

We consider the problem of error correction and detection for codes over large alphabets. Let $C \subseteq F_q^n$ be a linear code. The minimum distance of $C$, denoted $\delta(C)$, is defined as $\delta = \min_{x \neq y \in C} \delta(x, y)$, where $\delta(x, y)$ is the fractional Hamming distance between $x$ and $y$. Let $w$ be a word in $F_q^n$ and let $\delta(w, C) = \min_{x \in C} \delta(w, x)$ denote the distance of $w$ to $C$. $C$ is said to be $f(n)-$sparse for $t > 0$ if $|C| \leq f(n)$. $C$ is said to be $\epsilon$-biased if, for all $x \neq y \in C$, $1 - \frac{1}{q} - \epsilon \leq \delta(x, y) \leq 1 - \frac{1}{q} + \epsilon$.

We prove that sparsity and small bias are sufficient conditions to test membership of $w$ in $C$ or to find the closest codeword in $C$ to $w$, while querying only a constant number of symbols from $w$. Such codes are called *locally testable* and *self-correctable* codes. The following definitions are adapted to q-ary codes from [3].

DEFINITION 1.1. *Let $C \subseteq F_q^n$ be a linear code. $C$ is said to be strongly k-locally testable if there exists a constant $\epsilon > 0$ and a probabilistic algorithm $T$ called the tester, that given oracle access to a vector $v \in F_q^n$, queries the oracle at most $k$ times and accepts every $v \in C$ with probability 1 and rejects every $v \notin C$ with probability at least $\epsilon \cdot \delta(v, C)$.*

DEFINITION 1.2. *Let $C \subseteq F_q^n$ be a linear code. $C$ is said to be k-self correctable if there exist constants $\tau > 0$ and $0 < \epsilon < \frac{1}{2}$ and a probabilistic algorithm $SC$ called the self corrector, that given oracle access to a vector $v \in F_q^n$ that is $\tau-$ close to a codeword $c \in C$ and an index $i \in [n]$, queries the oracle at most $k$ times and computes $c_i$ with probability at least $1 - \epsilon$.*

Note that self-correctable codes include locally-decodable codes since self-correctability is independent of the encoder used and only uses the codewords themselves, while local decodability requires knowledge of the encoder and the messages.

## 1.1 Prior work and motivation

Dual-BCH codes are an important family of sparse unbiased codes and have a major role in Coding Theory. Proving local testability and self-correctability for these codes will add to their powerful properties, thus opening the door to new techniques in error detection and correction.
Moreover, local testability and self-correctability have many applications in Complexity Theory. Although codes satisfying these properties are not efficient (the rate tends to 0 as the block length tends to infinity), the local nature of the underlying algorithms proves to be very useful in constructing Probabilistically Checkable Proofs where one should accept correct proofs while only checking few locations in the proof.

Hadamard codes were the first codes shown to be locally testable and self-correctable in [1]. Recently, Kaufman and Litsyn proved in [2] that almost-orthogonal codes are locally testable and self-correctable. Such codes include dual-BCH codes. Kaufman and Sudan generalized these previous results and proved in [3] that sparse random binary linear codes are locally testable and decodable.
In [4], Kopparty and Saraf proved that random sparse binary linear codes are locally testable and local *list*-decodable. They strengthen the results of Kaufman and Sudan by correcting in the high-error regime, while using simpler proofs. They reduce the problem of testing/decoding codewords to that of testing/decoding linear functions under *distributions*. In [5], they prove that sparse low-bias codes over any abelian group are locally testable. Although their results subsume the results of [3] (except for removing the small bias property in the local testability case), we got our results independently and before their work was published.

## 1.2 Summary of results

We generalize the techniques in [3] to prove that q-ary sparse codes with small bias are locally testable and self-correctable. We follow the proof strategy of Kaufman and Sudan. We use properties of q-ary Krawtchouk polynomials and the McWilliams identity to bound the weight distributions of duals of sparse codes with small bias. These properties of q-ary Krawtchouk polynomials are non-obvious and were only obtained after the very recent work on zeros of discrete orthogonal polynomials in [6]. Thus, extending the results of [3] to q-ary codes requires a more detailed analysis of the underlying Krawtchouk polynomials and their properties.

Using the derived weight distribution bounds, we get the local testability result:

**Theorem 1.** *Let $F_q$ be the finite field of size $q$ and let $C \subset F_q$ be a linear code. For every $t \leq \infty$ and $\gamma > 0$, there exists a constant $k = k_{q,t,\gamma} \leq \infty$ s.t. if $C$ is $n^t$-sparse and $n^{-\gamma}$ biased, then $C$ is strongly $k-$locally testable.*

To get the self-correctability result, we apply the weight distribution bounds on punctured codes, where we removed one or two symbols from the codewords of the original codes. Namely, we prove:

**Theorem 2.** *For every $t \leq \infty$ and $\gamma > 0$, there exists a constant $k = k_{\tau,\gamma}$ such that if $C \subseteq F_q^n$ is a $n^t$-sparse and $n^{-\gamma}$-biased, then $C$ is $k$-self-correctable.*

## 1.3 Organization of the paper

In section 2, we derive properties of q-ary Krawtchouk polynomials and the McWilliams identity to bound the weight distributions of the dual code. In Section 3 and 4, we prove Theorems 1 and 2 respectively.

## 2 Weight distribution of duals of sparse codes

We use the McWilliams identity to relate the weight distribution of a code to that of its dual.

### 2.1 Q-ary Krawtchouk Polynomials and MacWilliams Identity

Let $q, k$ and $n$ be positive integers ($k, q \leq n$). Q-ary Krawtchouk polynomials $P_k(i, q, n)$ are orthogonal polynomials on $i = 0, \ldots, n$, with respect to the measure $\binom{n}{i}(q-1)^i$. They are defined as follows:

$$P_k(i, q, n) = \sum_{l=0}^{k} \binom{i}{l}\binom{n-i}{k-l}(-1)^l (q-1)^{k-l}$$

From now on, we will drop the $q$ and $n$ from the definition of $P_k$, when it's obvious.

#### 2.1.1 Properties

1. $P_k(0) = \binom{n}{k}(q-1)^k$.

2. For every $i$, $P_k(i, q, n) = P_k(n-i, \frac{q}{q-1}, n)(1-q)^k$.

3. $P_k(i)$ has $k$ real roots lying between $(1 - \frac{1}{q})n - k(1 - \frac{2}{q}) - \frac{2}{q}\sqrt{(q-1)k(n-k)}$ and $(1 - \frac{1}{q})n - k(1 - \frac{2}{q}) + \frac{2}{q}\sqrt{(q-1)k(n-k)}$.

4. Let $\mu_1 = (1 - \frac{1}{q})n - k(1 - \frac{2}{q}) - \frac{2}{q}\sqrt{(q-1)k(n-k)}$ and $\mu_2 = (1 - \frac{1}{q})n - k(1 - \frac{2}{q}) + \frac{2}{q}\sqrt{(q-1)k(n-k)}$.

    a. $P_k(i) \leq \frac{q^k}{k!}[(1 - \frac{1}{q})n - i]^k$, for all $i \in [n]$.
    
    b. $|P_k(i)| \leq \frac{q^k}{k!}[k + \frac{2}{q}(\sqrt{(q-1)k(n-k)} - k)]^k$, for $\mu_1 \leq i \leq \mu_2$.
    
    c. $P_k(i) \leq 0$, for $\mu_2 \leq i \leq n$ and odd $k$.

*Proof.* 1 and 2 follow from the definition of $P_k$. 3 is from Theorem 6 of [6]. 4 follows from 3 and basic manipulations of $P_k$. □

Let $C$ be a linear code of length $n$ over $F_q$. For every $i \in [n]$, let $B_i^C$ be the number of codewords in $C$ of weight $i$. The weight distribution of $C$ is given by the vector $< B_0^C = 1, \ldots, B_n^C >$. Let $C^\perp$ denote the dual code of $C$.

**Theorem 3** (MacWilliams Identity). *For a linear code over $F_q$ of length $n$, $B_k^{C^\perp} = \frac{1}{|C|}\sum_{i=0}^{n} B_i^C P_k(i, q, n)$, where $P_j(i, q, n)$ is the generalized Krawtchouk polynomial of degree $k$.*

The following proposition lists some properties of a linear code $C$.

**Proposition 4.** *[3] Let $C$ be an $n^t$−sparse linear code code over $F_q$ with $\delta(C) \geq 1 - \frac{1}{q} - n^{-\gamma}$, for some $t, \gamma > 0$. Then:*

- $B_0^C = 1$.

- $B_i^C = 0$ for all $i \in \{1, \ldots, (1 - \frac{1}{q})n - n^{1-\gamma}\}$.

- $\sum_{i=0}^{C} B_i^C \le n^t$.

- If $C$ is $n^{-\gamma}$ biased, then $B_i^C = 0$ for all $i \in \{n(1 - \frac{1}{q}) - n^{1-\gamma}, \ldots, n\}$.

The following is the equivalent of Claim 3.4 in [3] for $q$−ary linear codes.

**Claim 5.** *For every $\gamma > 0, c, t < \infty$, if $k \ge (t + c + 1)/\gamma$, then for any $n^t$−sparse set $S \subseteq F_q^n$, then:*

$$\left| \sum_{i=(1-1/q)n-n^{1-\gamma}}^{(1-1/q)n+n^{1-\gamma}} B_i^S P_k(i) \right| = o(n^{-c})P_k(0).$$

*Furthermore, if $k$ is odd, we have $\sum_{i=(1-1/q)n-n^{1-\gamma}}^{n} B_i^S P_k(i) = o(n^{-c})P_k(0)$.*

*Proof.*

$$\begin{aligned}
\left| \sum_{i=(1-1/q)n-n^{1-\gamma}}^{(1-1/q)n+n^{1-\gamma}} B_i^S P_k(i) \right| &\le \sum_{i=(1-1/q)n-n^{1-\gamma}}^{(1-1/q)n+n^{1-\gamma}} B_i^S |P_k(i)| \\
&\le max_i |P_k(i)| \sum_i B_i^S \\
&\le n^{(1-\gamma)k} \frac{q^k}{k!} |S| \\
&= o(n^{-c}) n^k \frac{q^k}{k!} \\
&= o(n^{-c}) P_k(0)
\end{aligned}$$

The third inequality follows from applying property 4.a of Krawtchouk polynomials for $(1-1/q)n - n^{1-\gamma} \le i \le (1 - 1/q)n + n^{1-\gamma}$. The first equality follows from the facts that $k \ge (t + c + 1)/\gamma$ and $|S| \le n^t$. The second part of the claim follows from the fact that $P_k(i) < 0$ for every $i \in \{(1-\frac{1}{q})n + n^{1-\gamma} \ldots n\}$. □

We will use the above claim to bound the weight enumerators of $C^\perp$. The following lemma is the equivalent of lemma 3.5 in [3].

**Lemma 6.** *Let $C$ be an $n^t$-sparse code in $F_q^n$ with $\delta(C) \ge 1 - \frac{1}{q} - n^{-\gamma}$. Then, for every $c, t, \gamma > 0$, there exists a $k_0$ s.t. for every odd $k \ge k_0$, $B_k^{C^\perp} \le \frac{P_k(0)}{|C|}(1 + o(n^{-c}))$. If $C$ is $n^{-\gamma}$-biased, then for every (odd and even) $k \ge k_0$, $B_k^{C^\perp} = \frac{P_k(0)}{|C|}(1 + \theta(n^{-c}))$.*

**Note** As mentioned in [3], the notation $f(n) = g(n) + \theta(n)$ means that, for every $\epsilon > 0$ and for large enough $n$, $g(n) - \epsilon h(n) \le f(n) \le g(n) + \epsilon h(n)$.

*Proof.* By MacWilliams Identity we get:

$$\begin{aligned}
B_k^{C^\perp} &= \frac{1}{|C|} \sum_{i=0}^{n} B_i^C P_k(i) \\
&= \frac{P_k(0)}{|C|} + \frac{1}{|C|} \sum_{i=1}^{n} B_i^C P_k(i) \\
&= \frac{P_k(0)}{|C|} + \frac{1}{|C|} \sum_{i=(1-\frac{1}{q})n-n^{1-\gamma}}^{n} B_i^C P_k(i), \\
&= \frac{P_k(0)}{|C|} + \frac{1}{|C|}(o(n^{-c})P_k(0)), \quad \text{using claim 1} \\
&= \frac{P_k(0)}{|C|}\left(1 + o(n^{-c})\right).
\end{aligned}$$

where the third equality follows from $\delta(C) \geq 1 - \frac{1}{q} - n^{-\gamma}$. The second part of the lemma, when $C$ is $n^{-\gamma}$-biased:

$$\begin{aligned}
B_k^{C^\perp} &= \frac{P_k(0)}{|C|} + \frac{1}{|C|} \sum_{i=1}^{n} B_i^C P_k(i) \\
&= \frac{P_k(0)}{|C|} + \frac{1}{|C|} \sum_{i=(1-\frac{1}{q})n-n^{1-\gamma}}^{(1-\frac{1}{q})n+n^{1-\gamma}} B_i^C P_k(i), \\
&= \frac{P_k(0)}{|C|} + \frac{1}{|C|}(\theta(n^{-c})P_k(0)), \quad \text{using claim 1} \\
&= \frac{P_k(0)}{|C|}\left(1 + \theta(n^{-c})\right).
\end{aligned}$$

where the second equality follows since $C$ is $n^{-\gamma}$-biased. □

## 3 Local Testing

We will use a canonical tester that uses codewords in the dual of a code to test membership of words in the code. The following tester is proposed in [3] for binary codes:

$T_k^v$ :

- Choose $y \in_U [C^\perp]_k$, where $[C^\perp]_k$ is the set of codewords in $C^\perp$ of weight $k$
- Accept if and only if $\langle y, v \rangle = 0$

If $v \in C$, $T_k^v$ accepts with probability 1. We want to estimate the probability that $T_k^v$ accepts $v$ when $v \notin C$. Following the same approach as in [3], we look at a new code that is the linear span of $C$ and $v$, $C||v = \bigcup_{\mu=0}^{q}(C + \mu v)$.

**Proposition 7.** *for $v \notin C$, $T_k^v$ rejects $v$ with probability $Rej_k(v) = 1 - \frac{B_k^{(C||v)^\perp}}{B_k^{C^\perp}}$.*

*Proof.* If $T_k^v$ accepts $v$, then $\langle y, v \rangle = 0$, then $y \in B_k^{(C||v)^\perp}$, since $\forall x \in C$ and $\mu \in F_q, \langle y, x \rangle = 0$ and $\langle y, x + \mu v \rangle = \langle y, x \rangle + \mu \langle y, v \rangle = 0$. If $y \in B_k^{(C||v)^\perp}$, then $\langle y, x + \mu v \rangle = 0, \forall x \in C, a \in F_q$. □

We now show that $Rej_k(v) = \Omega(\delta(v, C))$.

We will need the following two lemmas.

**Lemma 8.** *[3] For every $k$, for sufficiently large $n$ and for every $\tau < \frac{1}{2}$, $P_k(\tau n) \leq (1 - \tau)^k P_k(0)$.*

*Proof.* The proof is the same as in [3], where we use the fact that $P_k(0) = \binom{n}{k}(q-1)^k$ and property 4.a of $q$-ary Krawtchouk polynomials. □

The following lemma is the equivalent of Lemma 5.4 in [3].

**Lemma 9.** *Let $k, t, \gamma$ be constants. Let $\gamma' \leq \gamma/2$. For sufficiently large $n$, let $D$ be an $n^t$-sparse code in $F_q^n$ of distance at least $1 - \frac{1}{q} - n^{-\gamma}$. Let $\delta \leq \frac{1}{2}$ and let $a = max\{(1-\frac{1}{q})n - n^{1-\gamma'} - \delta n, \delta n\}$ and $b = (1-\frac{1}{q})n - n^{1-\gamma'}$. Then $\sum_{i=a}^{b} P_k(i) B_i^D \leq 2(q^2 + q) P_k(0) \cdot \min\left\{ (1 - \frac{q}{q-1}\delta)^{k-2}, \left(\frac{2q}{(q-1)}\delta\right)^{k-2} + \left(\frac{2q}{(q-1)} n^{-\gamma}\right)^{k-2} \right\}$.*

*Proof.* For a code with minimum distance $n(1 - \frac{1}{q}) - n^{1-\gamma}$, the Johnson bound states that, for $i \leq n(1 - \frac{1}{q}) - n^{1-\gamma/2}$, the number of codewords in a ball of radius $i$ is at most $\frac{qn^2}{(n - \frac{q}{q-1}i)^2}$.

Let $m_i = \frac{qn^2}{(n-\frac{q}{q-1}i)^2}$. Then, by the Johnson bound, $\sum_{j=0}^{i} B_i^D \leq m_i$, for all $i \leq b$. We get:

$$\sum_{i=a}^{b} P_k(i) B_i^D \leq \frac{(q-1)^k}{k!} \sum_{i=a}^{b} \left(n - \frac{q}{q-1}i\right)^k B_i^D$$

$$\leq \frac{(q-1)^k}{k!} \left(n - \frac{q}{q-1}i\right)^k m_a + \frac{(q-1)^k}{k!} \sum_{i=a+1}^{b} \left(n - \frac{q}{q-1}i\right)^k (m_i - m_{i-1})$$

Replacing $m_a$ by its value in the first term, we get

$$\frac{(q-1)^k}{k!}\left(n - \frac{q}{q-1}a\right)^k m_a \leq \frac{(q-1)^k}{k!}\left(n - \frac{q}{q-1}a\right)^k \times \frac{qn^2}{\left(n-\frac{q}{q-1}a\right)^2} \leq \frac{q(q-1)^k}{k!} n^2 \left(n - \frac{q}{q-1}a\right)^{k-2}$$

For the second term, note that $m_i - m_{i-1} \leq \frac{2q^2 n^2}{\left(n-\frac{q}{q-1}i\right)^3}$. Hence, we get:

$$\frac{(q-1)^k}{k!} \sum_{i=a+1}^{b} \left(n - \frac{q}{q-1}i\right)^k (m_i - m_{i-1}) \leq \frac{(q-1)^k}{k!} \sum_{i=a+1}^{b} \left(n - \frac{q}{q-1}i\right)^k \frac{2q^2 n^2}{\left(n - \frac{q}{q-1}i\right)^3}$$

$$\leq \frac{2q^2 n^2 (q-1)^k}{k!} \sum_{i=a+1}^{b} \left(n - \frac{q}{q-1}i\right)^{k-3}$$

$$\leq \frac{2q^2 n^2 (q-1)^k}{k!} (b-a) \left(n - \frac{q}{q-1}a\right)^{k-3}$$

$$\leq \frac{q^2 n^2 (q-1)^k}{k!} \left(n - \frac{q}{q-1}a\right)^{k-2}$$

Combining the above, we get

$$\sum_{i=a}^{b} P_k(i) B_i^D \leq \frac{(q^2+q) n^2 (q-1)^k}{k!} \left(n - \frac{q}{q-1}a\right)^{k-2}.$$

Substituting for the value of $a$ and using the bound $\frac{(q-1)^k n^k}{k!} \leq 2 P_k(0)$, we get:

$$\sum_{i=a}^{b} P_k(i) B_i^D \leq 2(q^2+q) P_k(0) \cdot \min\left\{\left(1 - \frac{q}{q-1}\delta\right)^{k-2}, \left(\frac{2q}{(q-1)}(\delta + n^{-\gamma})\right)^{k-2}\right\}.$$

Using the convexity of the function $f(x) = x^{k-2}$, we get:

$$\sum_{i=a}^{b} P_k(i) B_i^D \leq 2(q^2+q) P_k(0) \cdot \min\left\{(1 - \frac{q}{q-1}\delta)^{k-2}, \left(\frac{2q}{(q-1)}\delta\right)^{k-2} + \left(\frac{2q}{(q-1)}n^{-\gamma}\right)^{k-2}\right\}.$$

The lemma follows since for $x, y, z > 0$, $\min\{x, y+z\} \leq \min\{x, y\} + z$. □

Back to the weight enumerator of the dual code, we use the above two lemmas to prove the following lemma:

**Lemma 10.** *For every $c, t < \infty$ and $\gamma > 0$, there exists a $k_0$ such that if $C$ is an $n^t$-sparse code of distance $\delta(C) \geq 1 - \frac{1}{q} - n^{-\gamma}$ and $v \in F_q^n$ is $\delta$-far from $C$, then, for odd $k \geq k_0$,*

$$B_k^{(C||v)^\perp} \leq (1 - \frac{\delta}{2} + o(n^{-c})) P_k(0)$$

*Proof.* Let $\gamma' = \gamma/2$. We will prove the lemma for $k_0 = \max\{k_1, k_2, 16(q^2+q)\}$, where $k_1$ is chosen to be big enough so that Claim 5 applies and $k_2$ is the constant given by Lemma 6 as a function of $t, c$ and $\gamma$.

By the MacWilliams Identity, we have

$$B_k^{(C||v)^\perp} = \frac{1}{q|C|} \sum_{i=0}^n B_i^{(C||v)} P_k(i)$$
$$= \frac{1}{q|C|} \sum_{i=0}^n B_i^C P_k(i) + \frac{1}{q|C|} \sum_{i=0}^n B_i^{(C||v)} P_k(i)$$
$$= \frac{1}{q} B_k^{C^\perp} + \frac{1}{q|C|} \sum_{i=0}^n B_i^{(C||v)} P_k(i)$$

By Lemma 6, we have $B_k^{C^\perp} \leq \frac{P_k(0)}{|C|}(1 + o(n^{-c}))$. Hence, it's enough to prove

$$\frac{1}{q|C|} \sum_{i=0}^n B_i^{(C||v)} P_k(i) \leq (1 - \delta + o(n^{-c})) \frac{P_k(0)}{|C|}.$$

Applying Claim 5 to $(C + \mu v)$, we get $\sum_{i=n(1-\frac{1}{q})-n^{1-\gamma'}}^n B_i^{(C||v)} P_k(i) = o(n^{-c}) \cdot P_k(0)$. Now, it suffices to prove

$$\sum_{i=0}^{n(1-\frac{1}{q})-n^{1-\gamma'}} B_i^{(C||v)} P_k(i) = (1 - \delta + o(n^{-c})) P_k(0).$$

Since $\delta(C) \geq 1 - \frac{1}{q} - n^{-\gamma}$, $B_i^{C+\mu v} = 0$ for every $i = \{0, \ldots, n(1-\frac{1}{q}) - n^{1-\gamma'} - \delta n\}$, except possibly for $i = \delta n$ ($v \in C + v$ and $\delta(v, C) \leq \delta$). If $\delta n \leq n(1-\frac{1}{q}) - n^{1-\gamma'}$, $B_{\delta n}^{C+av} = 1$. Thus, we have

$$\sum_{i=0}^{n(1-\frac{1}{q})-n^{1-\gamma'}} B_i^{C+\mu v} P_k(i) \leq P_k(\delta n) + \sum_{i=a}^b B_i^{C+\mu v} P_k(i),$$

where $a = \max\{n(1-\frac{1}{q}) - n^{1-\gamma'} - \delta n, \delta n\}$ and $b = n(1-\frac{1}{q}) - n^{1-\gamma'}$. Using the bound in 9 for these values of $a$ and $b$, we get:

$$\sum_{i=0}^{n(1-\frac{1}{q})-n^{1-\gamma'}} B_i^{C+\mu v} P_k(i) \leq P_k(0)\Big((1-\delta)^k + 2(q^2+q)\min\Big\{(1 - \tfrac{q^2}{(q-1)^2}\delta)^{k-2}, \Big(\tfrac{2q}{q-1}\delta\Big)^{k-2}\Big\} + O(n^{-\gamma'(k-2)})\Big).$$

Now, we want to prove that $(1-\delta)^k + 2(q^2+q)\min\Big\{(1 - \tfrac{q}{q-1}\delta)^{k-2}, \Big(\tfrac{2q}{(q-1)}\delta\Big)^{k-2}\Big\} \leq (1-\delta)$, for $0 \leq \delta < 1/2$.

- For $\delta \leq \frac{q-1}{8q^2(q+1)}$ and $k \geq 2 + \frac{2q}{q-1}$, we have
  $(1-\delta)^k + 2(q^2+q)\Big(\tfrac{2q}{(q-1)}\delta\Big)^{k-2} \leq (1-\delta)^k + 2(q^2+q)\Big(\tfrac{2q}{(q-1)}\delta\Big)^2 \leq 1 - \tfrac{k}{2}\delta + \tfrac{q}{q-1}\delta \leq 1 - \delta$.

- For $\frac{q-1}{8q^2(q+1)} \leq \delta \leq 1/2$ and $k \geq 16(q^2+q)$, we have
  $(1-\delta)^k \leq \tfrac{1}{4}$ and $2(q^2+q)(1-\tfrac{q}{q-1}\delta)^{k-2} \leq \tfrac{1}{4}$. Hence $(1-\delta)^k + 2(q^2+q)(1-\tfrac{q}{q-1}\delta)^{k-2} \leq \tfrac{1}{2} \leq 1 - \delta$.

The lemma follows since $k \geq \frac{t+c+1}{\gamma}$ as in Claim 5. □

Finally, we prove the main theorem for local testability for small-bias codes over large alphabets.

**Proof of Theorem 1** Given $t$ and $\gamma$, let $k$ be an odd integer greater than $k_0$ as given by Lemma 10. The rest of the proof is the same as the proof of Theorem 5.5 in [3].

## 4 Self-correctability

Below is the canonical self corrector for the code $C$ that uses the dual code $C^\perp$. This algorithm is used in [3] for binary codes.

$SC_k^v(i)$:

- Choose $y \in_U [C^\perp]_{k,i}$, where $[C^\perp]_{k,i}$ is the set of codewords in $C^\perp$ of weight $k$ that have a non-zero value at index $i$.

- Compute $(-y_i)^{-1} \sum_{\{j \in [n]-\{i\} s.t. y_j \neq 0\}} v_j$.

$SC_k^v(i)$ have oracle access to $v$ such that $\delta(v, c) < \frac{1}{2k}$, for every $c \in C$. It makes $k-1$ queries to $v$. We want to estimate the probability that $SC_k^v(i)$ doesn't compute $c_i$.

We'll start by estimating the probability that $y \in_U [C^\perp]_k$ has non-zero entries at indices $i$ and $j$. As in [3], let $C_i = \{\pi_{-i}(c), c \in C\}$, where $\pi_{-i}(c_1, \ldots, c_{i-1}, c_i, c_{i+1}, \ldots, c_n) = (c_1, \ldots, c_{i-1}, c_{i+1}, \ldots, c_n)$.

**Proposition 11.** *[3]* Let $\pi_{-i}^{-1}(c_1, \ldots, c_{i-1}, c_{i+1}, \ldots, c_n) = (c_1, \ldots, c_{i-1}, 0, c_{i+1}, \ldots, c_n)$ and $\pi_{-i}^{-1}(S) = \{\pi_{-i}^{-1}(y) | y \in S\}$. Then $[C^\perp]_{k,i} = [C^\perp]_k - \pi_{-i}^{-1}\left([(C^{-i})^\perp]_k\right)$ and hence, $|[C^\perp]_{k,i}| = |[C^\perp]_k| - |[(C^{-i})^\perp]_k|$.

*Proof.* If $y \in C^\perp$ with $y_i = 0$, then $\pi_{-i}(y) \in (C^{-i})^\perp$ and hence $\{\pi_{-i}(y) | y \in C^\perp | y_i = 0\} \subseteq (C^{-i})^\perp$. We'll show that $(C^{-i})^\perp = \{\pi_{-i}(y) | y \in C^\perp | y_i = 0\}$.
$(C^{-i})^\perp$ is the dual of $C^{-i}$, hence $|(C^{-i})^\perp| = \frac{q^{n-1}}{|C^{-i}|} = \frac{q^{n-1}}{|C|}$ since $\delta(C) \geq \frac{2}{n}$. Thus, $|(C^{-i})^\perp| = \frac{1}{q}|C^\perp|$.
On the other hand, $|\{\pi_{-i}(y) | y \in C^\perp | y_i = 0\}| \geq \frac{1}{q}|C^\perp|$. Therefore $[C^\perp]_{k,i} = [C^\perp]_k - \pi_{-i}^{-1}([(C^{-i})^\perp]_k)$. □

Extending the puncturing to two indices $i \neq j$, let $C^{-\{i,j\}}$ be the projection of $C$ on $[n] - \{i, j\}$ and let $\pi_{-\{i,j\}}$ be that projection. Thus, we get:

**Proposition 12.** *[3]* For every $i \neq j$, $[C^\perp]_{k,\{i,j\}} = [C^\perp]_k - \pi_{-i}^{-1}([(C^{-i})^\perp]_k) - \pi_{-j}^{-1}([(C^{-j})^\perp]_k) + \pi_{-\{i,j\}}^{-1}([(C^{-\{i,j\}})^\perp]_k)$. Hence, $|[C^\perp]_{k,\{i,j\}}| = |[C^\perp]_k| - |[(C^{-i})^\perp]_k| - |[(C^{-j})^\perp]_k| + |[(C^{-\{i,j\}})^\perp]_k|$.

*Proof.* Same as in [3]. □

Using what we know about weight distributions of the above dual codes, we derive the probability that $y_j \neq 0$ when $y$ is chosen at random from $[C^\perp]_{k,i}$.

**Lemma 13.** *For every $\gamma > 0$ and $c, t < \infty$, there exists $k$ such that for sufficiently large $n$, if $C \subseteq F_q^n$ is an $n^t$-sparse, $n^{-\gamma}$-biased linear code, then for every $i \neq j \in [n]$, the probability that $y_j \neq 0$, when $y$ is chosen at random from $[C^\perp]_{k,i}$ is $\frac{k-1}{n-1} + \theta(n^{-c})$.*

*Proof.* $\Pr_{y \in_U [C^\perp]_{k,i}}[y_j \neq 0] = \frac{|[C^\perp]_{k,\{i,j\}}|}{|[C^\perp]_{k,i}|}$. Using the above two propositions, we can calculate those two quantities. $C$, $C^{-i}$, $C^{-j}$ and $C^{-\{i,j\}}$ are $n^t$-sparse, $n^{-\gamma}$-biased codes, with respective block lengths $n$, $n-1$, $n-1$ and $n-2$. Note also that they all have the same size. Picking $k$ large enough to apply Lemma 1 to these codes, we get:

$$|[C^\perp]_k| = \frac{(q-1)^k}{|C|} \left( \binom{n}{k} + \theta(n^{k-c}) \right)$$

$$|[(C^{-i})^\perp]_k| = \frac{(q-1)^k}{|C|} \left( \binom{n-1}{k} + \theta(n^{k-c}) \right)$$

$$|[(C^{-j})^\perp]_k| = \frac{(q-1)^k}{|C|} \left( \binom{n-1}{k} + \theta(n^{k-c}) \right)$$

$$|[(C^{-\{i,j\}})^\perp]_k| = \frac{(q-1)^k}{|C|} \left( \binom{n-2}{k} + \theta(n^{k-c}) \right)$$

Now we use Proposition 12 to obtain:

$$|[C^\perp]_{k,i}| = \frac{(q-1)^k}{|C|} \left( \binom{n-1}{k-1} + \theta(n^{k-c}) \right)$$

and

$$|[C^\perp]_{k,\{i,j\}}| = \frac{(q-1)^k}{|C|} \left( \binom{n-2}{k-1} \frac{k-1}{n-k} + \theta(n^{k-c}) \right).$$

Finally,

$$\frac{|[C^\perp]_{k,\{i,j\}}|}{|[C^\perp]_{k,i}|} = \frac{k-1}{n-1} + \theta(n^{-c}).$$

□

Now, we prove the main lemma that bounds the error probability of $SC_k^v(i)$.

**Lemma 14.** *[3] For every $t < \infty$, $\gamma > 0$, there exists a $k = k_{t,\gamma} < \infty$ such that: if $C$ is an $n^t$-sparse, $n^{-\gamma}$-biased linear code in $F_q^n$ and $v \in F_q^n$ is $\tau$-close to $C$, then for every $i \in [n]$, $\Pr[SC_k^v(i) \neq c_i] \leq k\tau + \theta(1/n)$.*

*Proof.* Pick $k$ large enough to apply Lemma 13 with $c = 2$. Hence, for $y \in_U [C^\perp]_{k,i}$ and $i \neq j$, $\Pr[y_j \neq 0] = \frac{k-1}{n-1} + \theta(n^{-2})$.
Let $E$ be the set of errors in $v$, i.e. $E = \{j \in [n] | v_j \neq c_j\}$. Since $v$ is $\tau$-close to $C$, then $|E| \leq \tau n$. Let $S_y$ be the set of non-zero symbols in $y$, i.e. $S_y = \{j \in [n] | y_j \neq 0\}$. $SC_k^v(i)$ will err only if the errors on $v$ line up with the non-zero symbols of $y$, and hence only if $E \cap S_y \neq \phi$. Therefore, $\Pr[SC_k^v(i) \neq c_i] = \Pr[E \cap S_y \neq \phi] \leq |E| \max_{j \in E} \Pr_y[y_j \neq 0] \leq k\tau + \theta(n^{-1})$. □

Since all the strings $v$ we are considering are at distance less than $\frac{1}{2k}$, we get the result of Theorem 2 for the self-correctibilty of sparse small-biased linear codes in $F_q^n$.

# 5 Conclusion and future work

We proved that sparse codes with small bias over large alphabets are locally testable and self-correctable. We used properties of the generalized Krawtchouk polynomials and some basic results from coding theory like the McWilliams identity and the Johnson bound. The next step is to relax the small bias condition, while maintaining a good minimum distance. Kaufman and Sudan were able to remove the small bias condition for local testability in the case sparse binary codes with large distance in [3]. Even in the binary case, removing the small bias condition for self-correctability of sparse codes with large distance is still an open problem. Moreover, the techniques used in [4] might be extended to get *list*-decodability of sparse q-ary codes.

# Acknowledgments

The author would like to thank Venkatesan Guruswami and Paul Beame for their guidance and helpful discussions and the anonymous referees for valuable comments.